\documentstyle[12pt]{article}

\def\Journal#1#2#3#4{{#1} {\bf #2}, #3 (#4)}

\def\NPB{{\em Nucl. Phys.} B}
\def\PLB{{\em Phys. Lett.}  B}
\def\PRL{\em Phys. Rev. Lett.}
\def\PR{\em Phys. Rept.}
\def\PRD{{\em Phys. Rev.} D}

\def\SJNP{\em Sov. J. Nucl. Phys.}

\title{
\begin{flushright}
{\large Yaroslavl State University\\
        Preprint YARU-HE-97/09 \\
        hep-ph/9712423} \\[12mm]
\end{flushright}
       {\LARGE\bf Field-induced axion decay $a \to e^+ e^-$} \\ 
       {\LARGE\bf in KSVZ-model}} 

\author{{\Large\bf N.V.~Mikheev, O.S.~Ovchinnikov} \\[2mm]
        {\large\it Yaroslavl P.G.~Demidov State University, 
                   Yaroslavl 150000, Russia} \\
        {\large\it E-mail: \quad 
                   mikheev@yars.free.net \quad 
                   ovoles@univ.uniyar.ac.ru} \\[4mm]
        {\Large\bf and} \\[4mm]
        {\Large\bf L.A.~Vassilevskaya} \\[2mm]
        {\large\it Moscow M.V.~Lomonosov State University, 
                   V-952, Moscow 117234, Russia} \\
        {\large\it E-mail: \quad vasilevs@vitep5.itep.ru}}

\date{}

\begin{document}

\maketitle

\bigskip

\begin{abstract}
{\normalsize 
The axion decay into electron-positron pair $a \to e^+ e^-$ 
is studied in an external magnetic field in 
KSVZ-model where axions have only induced coupling to leptons. 
The axion lifetime in the field decreases with energy and field 
strength to seconds. 
}
\end{abstract}

\vfill 

\begin{center} 
{\large\it Talk given at the International Workshop} \\ 
{\large\it on Particle Physics and the Early Universe ``COSMO'97'',} \\ 
{\large\it Ambleside, Lake District, England, 15-19 September 1997.} 
\end{center} 

\thispagestyle{empty} 

\newpage 

\large 

Peccei-Quinn ($PQ$) symmetry $U_{PQ}(1)$~\cite{P1}, with its 
accompanying axion~\cite{WW}, continues to be an attractive solution 
to the strong CP problem in QCD. At present axions are of great interest 
not only in theoretical aspects of elementary  particle physics, but in 
some astrophysical and cosmological applications as  
well~\cite{Tur,Raf1,Raf2}. 
Although the original axion is excluded experimentally, modified 
$PQ$ models with very light and very weakly coupled, so called, invisible 
axions are still tenable (see, for example the recent review~\cite{P2}). 
Invisible axion models are classified into two types 
depending on whether or not they have direct couplings to leptons.  The, 
so called, KSVZ axions~\cite{KSVZ} are hadronic axions with only induced 
coupling to leptons.  The, so called, DFSZ axions~\cite{DFSZ}, 
arise in models where axions couple to leptons already at tree level. 
 
In this paper we analyze the influence of an external magnetic field 
on the axion decay into electron-positron pair via a photon intermediate 
state $a \to \gamma \to e^+ e^-$ in KSVZ model.  
The lagrangian describing the effective axion-photon coupling is: 
 
\begin{equation} 
{\cal L}_{a \gamma} =  g_{a \gamma} \; \partial_{\mu} \, A_\nu \; 
\tilde F_{\nu\mu} \; a, 
\label{eq:Lag} 
\end{equation} 
 
\noindent 
$g_{a \gamma}$ is a constant with the dimension $(energy)^{-1}$; 
$A_\mu$ is the 
4-potential of the quantized electromagnetic field, 
$\tilde F$ is the dual external field tensor. In the second order of 
the perturbation theory a matrix element can be presented in the form: 
 
\begin{eqnarray} 
S & = & - \frac{ i e \; g_{a \gamma}}{\sqrt{2 E_a V}} \, 
 \int d^4 x \, \left ( {\bar \psi}(x) \, \hat h \, \psi(x) \right ) 
\, e^{-i q x}, 
\label{eq:S1} \\ 
h_{\alpha} & = & (q \tilde F G (q))_\alpha = 
q_\mu \tilde F_{\mu\nu} G (q)_{\nu\alpha}, 
\nonumber 
\end{eqnarray} 
 
\noindent where $e > 0$ is the elementary charge; $\psi(x)$ is the known 
solution of the Dirac equation in a magnetic field; $q_\alpha$ is the 
4-momentum of the decaying axion; $G_{\alpha \beta}$ is the photon 
propagator in the field. It is convenient to use the expression 
for the propagator in a diagonal form~\cite{Shab}: 
 
\begin{eqnarray} 
G_{\alpha \beta} =  \sum_{\lambda=1}^3 \; 
\frac { b_\alpha^{(\lambda)} b_\beta^{(\lambda)} }{ (b^{(\lambda)})^2 } \; 
\frac{ - i }{ q^2 - \ae^{(\lambda)} }, \;\;\;\; 
b_\alpha^{(\lambda)} b_\alpha^{(\lambda')} = 
\delta_{\lambda \lambda'} \; 
(b_{\alpha}^{(\lambda)})^2. 
\label{eq:G} 
\end{eqnarray} 
 
\noindent Here the basis vectors $b_\alpha^{(\lambda)}$ are the  
eigenvectors of the photon polarization tensor, $\ae^{(\lambda)}$ 
are the eigenvalues. Notice that only the basis vector 
$b_\alpha^{(2)} = (q \tilde F)_\alpha$ gives a contribution to the decay. 
 
Here we present the results of our calculations for the axion lifetime 
in the limiting case $E_a^2 \sin^2 \theta \gg e B $ ($E_a$ is the energy 
of the decaying axion; $\theta$ is the angle between the vectors of the 
magnetic field strength ${\vec B}$ and the momentum of the axion ${\vec q}$), 
when electron and positron are born in the states corresponding to the 
highest Landau levels. In this case the eigenvalue $\ae^{(2)}$ is described 
by the expression: 

\begin{eqnarray} 
\ae^{(2)} 
\simeq \frac{ 9 \cdot 3^{1/6} \, \Gamma^4( {2 \over 3} ) }{ 14 \pi^2 } \; 
\alpha \; (1 - i \sqrt{3}) \, (e^2 qFFq)^{1/3}. 
\label{eq:Ae} 
\end{eqnarray} 
 
\noindent To obtain the decay probability one has to carry out 
a non-trivial integration over the phase space of the electron-positron 
pair taking their specific kinematics in the magnetic field into account. 
The result can be presented in the form: 
 
\begin{eqnarray} 
W  \simeq 2,46 \cdot 10^3 \; g^2_{a \gamma} \; (eB)^{4/3} \; 
E_a^{1/3} \; \sin^{4/3} \theta . 
\nonumber 
\end{eqnarray} 
 
\noindent The corresponding expression for the axion lifetime is: 
 
\begin{eqnarray} 
\tau^{KSVZ} \simeq 1,16 \; 
\left ( \frac{10^{-10}}{g_{a \gamma} \, GeV} \right )^2 \; 
 \left ( \frac{E_a}{10 \, MeV} \right )^{-1/3} \;  
\left ( \frac{10^{15} G}{B \sin \theta} \right )^{4/3} \; s. 
\label{eq:T1} 
\end{eqnarray} 
 
For comparison let us present here the expression for the axion 
lifetime in vacuum~\cite{Raf1}: 
 
\begin{equation} 
\tau^{(0)} (a \to 2\gamma) \sim 1,3 \cdot 10^{44} \; 
\left ( \frac{ 10^{-10} }{ g_{a \gamma} \; GeV } \right )^2 \; 
\left ( {10^{-3} \; eV \over m_a} \right )^4 \; 
\left ( {E_a \over 10 \, MeV} \right ) \; s. 
\label{eq:T0} 
\end{equation} 
 
\vspace{5mm} 
 
The work was supported by Grant INTAS 96-0659.

\end{document}